\documentclass[12pt]{article}
\usepackage{epsfig}
\usepackage{amsfonts}
\usepackage{amsopn}
\usepackage{amsmath}

\usepackage{multirow}

\DeclareMathOperator\erf{erf}
\DeclareMathOperator\erfi{erfi}

\title{Relativistic Black-Scholes model}

\author{
Maciej Trzetrzelewski \thanks{e-mail: maciej.trzetrzelewski@gmail.com} \\ \\
 M. Smoluchowski Institute of Physics, \\
Jagiellonian University, \\
\L{}ojasiewicza, St. 11, 30-348 Krak\'ow, \\
Poland}

\begin{document}
\date{}
\maketitle

\abstract{Black-Scholes equation, after a certain coordinate transformation, is equivalent to the heat equation. On the other hand the relativistic extension of the latter, the telegraphers equation, can be derived from the Euclidean version of the Dirac equation. Therefore the relativistic extension of the Black-Scholes model follows from relativistic quantum mechanics quite naturally. 
We investigate this particular model for the case of European vanilla options. Due to the notion of locality incorporated in this way, one finds that the  
 volatility frown-like effect appears when comparing to the original Black-Scholes model.
}

\section{Introduction}

Among many unrealistic assumptions made in the Black-Scholes model \cite{BS}, one is particularly problematic - constant volatility $\sigma$. When the current market data are used against the Black-Scholes formula one finds that $\sigma$ must in fact depend on the strike $K$, and time to expiry $T$, in order to make the pricing formula work. Therefore the market data imply that $\sigma$ is not constant but a function $\sigma_I(K,T)$ - called implied volatility. 
The curve $\sigma_I(K,T)$ with $T$ fixed, is often U shaped so that it became a standard practice to call it a volatility smile. However that shape can also look more like a skew (a smirk) or a frown depending on the data/market one is considering.

Clearly, the fact that $\sigma_I(K,T)$ is not constant falsifies the Black-Scholes model. However, it is also well known that this situation was completely different  before the market crash in late 80'. In the equity market before 1987, the implied volatility was indeed fairly constant - why it is not constant nowadays \cite{Derman} ?

One could explain this problem by blaming everything on yet another unrealistic assumption of the Black-Scholes model - that the underlier $S_t$ undergoes the geometric Brownian motion
\begin{equation} \label{gbm}
dS_t/S_t = \mu dt + \sigma dW_t,  \ \ \ \ \mu\in \mathbb{R}, \ \ \sigma >0
\end{equation}
(where $W_t$ is a Wiener process). It follows form (\ref{gbm}) that log-returns (i.e. returns of $\ln S_t$) have Gaussian distribution. However it is very well known \cite{Mandelbrot} that the actual log-returns  are not distributed like that - instead they exhibit fat tails. Therefore a rather natural way to generalize (\ref{gbm}) is to replace $W_t$ with the process whose PDF exhibits fat tails corresponding to the ones observed in the markets. 
However a careful inspection shows that this cannot be the main reason of the volatility smile observed today. The point is that even before 1987 the log-return distribution revealed fat tails (note that Mandelbrot's paper \cite{Mandelbrot} was published in 1963) but at the same time the  Black-Scholes model was working well. This is clearly an issue. If fat tails are the reason of all these discrepancies then why the constant volatility assumption was correct before 1987?

Because of practical reasons, models that consider generalizations of $W_t$ are not very popular and the development in this subject went in a completely different direction. Instead of changing $W_t$, financial practitioners  prefer to leave $W_t$ unchanged and assume that $\sigma$  is a function $\sigma=\sigma(S,t)$ - called local volatility \cite{localvol}. Then the smile is explained by assuming that  $\sigma$ increases for large $|\ln S_t|$ - if this is the case then the tails of the Gaussian distribution will become fatter.  There is a way to find $\sigma(S,t)$ directly using the market data  \cite{Dupire}. However it turns out that this model also has its drawbacks i.e. while the smile can be accommodated, its dynamics (the dynamics of the smile when the strike changes) is not captured correctly. This brings us to further generalization by assuming that $\sigma$ itself is a stochastic process \cite{Heston} 
\begin{equation}  \label{svol}
d\sigma_t = \alpha(\sigma_t,t)dt + \beta(\sigma_t,t) d\tilde{W}_t
\end{equation}
(here $\alpha$ and $\beta$ are some deterministic functions, $\tilde{W}_t$ is another Wiener process, different from $W_t$ in (\ref{gbm}) ). This generalization is counter intuitive: the amplitude $\sigma$, that multiplies the random factor $dW_t$, is stochastic now, but shouldn't $dW_t$ contain all the randomness? Moreover, stochastic volatility models also fail in certain situations e.g. in the limit $T\to 0$ where $T$ is the time to maturity \cite{Gatheral}. This could be a  motivation to generalize further and introduce jumps i.e. discontinuous moves of the underling $S_t$ \cite{jumps}.\footnote{The reader will note that the line of reasoning presented here differs from the chronological way these ideas were considered. Jumps were introduced in 1976, three years after the Black-Scholes paper,  stochastic volatility in 1993, local volatility in 1994.}

It is clear that this way of making the models more general is likely to have little explanation power. These models may fit very well to the market data but in say 10 years from now they will most probably fail in some situations and one will have to make some other generalizations to fit the market data again. This implies that the stochastic volatility models are non falsifiable.

For example, if we agree on the fact that volatility $\sigma$ is a stochastic process and satisfies (\ref{svol}) then there is a priori no reason not to go further and assume that $\beta$ is also stochastic. This would make  our model even better calibrated to the market data. The possibilities are quite frankly unlimited and if it weren't for the fact that Monte Carlo simulations are time consuming, they would certainly be investigated. Because one can always augment the model in such way that it will be consistent with the data, it follows that the model cannot be falsified.

Nevertheless most financial practitioners prefer stochastic volatility models because then, one can still use Ito calculus and obtain some analytical, robust results (otherwise, when $dW_t$ is not a Wiener process, little exact results/methods are known \cite{nonWiener}). It may seem unusual, from the scientific point of view, that robustness of the model is used as a criteria of its applicability. However quantitative finance, unlike Physics, is not about predicting future events but about pricing financial instruments today. Therefore as long as our models are calibrated to the market, minimize arbitrage opportunities and are stable against small fluctuations of the data, there is a priori no problem in the existence of plethora of possible models in this subject. 

In Physics the situation is much different. There, we care about predictions and recalibration is not allowed. A theory that contains parameters and degrees of freedom in such amount that can explain any experimental data, by just appropriately fitting them, cannot be falsified and hence is physically useless\footnote{At this point it is worth noting that in theoretical physics there are constructions (such as string theory) which suffer from making no predictions in this sense.}. For every theory, it is absolutely crucial to have an example of an experiment which outcome may, in principle, disagree with the results of the theory. This way of thinking is in fact opposite to the way one proceeds in finance.

In this paper we would like to approach these issues from a different perspective. It is well known that algorithmic trading became more and more popular in the 80' - increasing the changes of the prices, per second. However there exists a  concrete underlying limitation for market movements: the change of any price $S(t)$ cannot be arbitrary large per  unit of time i.e. there exist maximal speed  $c_M$ such  that $\dot{S}(t) < c_M$ (market speed of light, $[c_M]=s^{-1}$).  An obvious proof of this assertion comes from the fact that the speed of information exchange is limited by the speed of light. It seems that this limitation should not be very restrictive since light travels about $30$cm per nano-second($ns$). Assuming that servers of two counter parties are, say, $30$cm from each other, it takes at least $1ns$ to send an order.  Therefore we should not see any relativistic effects, unless we are considering situation in which there are at least billions ($10^9$) orders per second, sent to a single server. At this point it is clear that future development of high frequency trading may in principle influence the situation considerably. In fact it has already been observed that spatial separation of trading counter parties is a potential source of statistical arbitrage, see e.g. \cite{statarb}. 

 However there is one feature of every liquid market whose consequences are seen already and hence we would like to discuss it in more details. Any price $S(t)$ going (say) up from $S(t)$ to $S(t+\Delta t)>S(t)$, must overcome all the offers made in the interval $[S(t),S(t+\Delta t)]$. This introduces a natural concept of friction/resistance in the markets simply because there is always somebody who thinks that the price is too high. This situation is similar to what happens in physical systems e.g. electrons in conductors. An electron can a priori move with arbitrary (but less than $c$ - the speed of light) velocity. However due to constant collisions with atoms of the conductor the maximal velocity is in fact bounded even more. The drift velocity of electrons can be as small as e.g. $1m/h$. Perhaps a better physical example is light traveling in a dense media  where the effective speed of light is $c/n$ where $n$ is the refractive index (e.g. $n=1.3, \ \  1.5,  \ \ 2.4$ for water, glass and diamond respectively). In extreme situations, when light travels through the Bose-Einstein condensate, the effective speed of light can be as small as  $1m/s$ \cite{condensate}. Another good examaple is given by graphene surfaces (for a review see e.g. \cite{graphene}) for which description of electrons is effectively given by the massless Dirac equation. 
 
To see that this resistance effect is big in the markets let us consider the logarithm $x(t)=\ln S(t)$ and the corresponding bound on the derivative of $x(t)$, $|\dot{x}(t)| = |\dot{S}|/S < c_M/S$. If we assume that the order of the underlying is about 100$\$$ and that $c_M$ is at least $10^9 s^{-1}$ then we obtain $|\dot{x}|<10^7s^{-1}$. On a daily basis this implies that the difference $\Delta x:=|x(\hbox{day}) - x(\hbox{previous   day})|$
can a priori be as big as $10^7 \cdot 3600 \cdot 24 = 8.64 \cdot 10^{12}$.  However at the same time nothing alike is observed in the market.  The value of $\Delta x$ for any asset was, to our knowledge, never bigger than $1$. We have analyzed top 100 companies (considering their market capitalization as of March 2012) of the SP500 index. We order them w.r.t. decreasing maximal  absolute value of their log-returns.  The list of first 5 of them is presented below
 
 {\small \hspace{-0.6cm}
\begin{tabular}{l*{6}{l}r}
Company              & log-return & market move (close) & date  \\
\hline
WMT &  -0.735707  & 0.0192 $\to$ 0.0092 & Dec 1974   \\
AAPL &   -0.730867 & 26.18 $\to$ 12.60 & Sep 2000   \\
INTC &  0.698627  & 0.0091 $\to$  0.0183 & Jan 1972   \\
C &  -0.494691    & 24.53 $\to$  14.96 & Feb 2009   \\
ORCL &  -0.382345   & 0.61 $\to$  0.42 & Mar 1990    
\end{tabular} }

We see that the magnitude of log-returns may be of order of $10^0$, not $10^{12}$. This implies that there is a huge resistance in the market for the price to move up or down. Therefore one may conclude that the effective maximal velocity of $S(t)$ is much smaller than $c_M$. For completeness we also performed the same analysis for other markets like Forex majors, precious metals and major indices and found that all the log-returns are small. This confirms our claim that the maximal value of $|\dot{x}|$ is smaller than $1$ per day. We will use the notation $c_m$ for the upper bound of $|\dot{x}|$.

In the next section we  present a basic idea investigated in this paper -  the existence of the bound on log-returns implies that the corresponding PDF, $p(x,t)$, cannot be positive everywhere but must be $0$  for $|x| > x_{max}:=c_m t$.  This generically introduces a skew/smirk of the volatility when comparing to the Gaussian distribution. Based on the market data analysed above we claim that this effect can in fact be noticeable.
The main question is then, in what way we can generalise the Black-Scholes model so that the finiteness of $c_m$ is taken into account. Towards this direction it seems natural we study the relativistic generalisation of the diffusion equation - this idea is of course not new, see \cite{review} for a comprehensive review, for more recent works on relativistic extensions of pricing equations see \cite{relquant}.

The paper is organised as follows. In Section 2 we present the main idea investigated in the paper.
In Section 3 we review the correspondence between the relativistic diffusion equation, the telegraphers equation and the Dirac equation found a few decades ago \cite{Goldstein, Kac, GJKS, JS, Orsingher1,Orsingher2}. Based on the discussion in Section 3, in Section 4 we propose a generalisation of the geometric Brownian motion to the case where finite velocity $c_m$ is taken into account. We then  use the pricing formula for vanilla derivatives and   
perform the   $1/c_m$ expansion for option prices. This result can  then be used to evaluate the implied volatility exactly, when $c_m$ is large.

\section{Main idea}

Consider a model that takes into account finite maximal speed of propagation of information (locality in the market). The speed of $S(t)$ and hence $x=\ln S(t)$ is bounded.  Let $p(x,t)$ be the corresponding probability density and let us  expand it  about the normal distribution as follows
\begin{equation} \label{expansion}
p(x,t) = \frac{e^{-\frac{x^2}{2\sigma^2 t}}}{\sqrt{2\pi \sigma^2 t}}\left(1+\frac{1}{c_m^2}f(x,t,\sigma)+\ldots \right),
\end{equation}
where $\sigma$ is the volatility in the Black-Scholes model and where $f(x,t,\sigma)$ is of compact support, corresponding to the $1/c_m^2$ corrections of this expansion (anticipating results from Section 4, we do not consider $1/c_m$ corrections). Note that $f(x,t,\sigma)$ must be such that the distribution $p(x,t)$ is $0$ for $|x|\ge x_{max}:=c_m t$ (i.e. $f$ is $\frown$ shaped) - a result following simply from locality.  

We are interested in the $x$, and $t$ dependent volatility $\sigma_{DI}(x,t,\sigma)$ (density-implied volatility) so that
\begin{equation}   \label{expansion1}
p(x,t)= \frac{e^{-\frac{x^2}{2\sigma^2_{DI} t}}}{\sqrt{2\pi \sigma_{DI}^2 t}}.
\end{equation}
Density implied volatility $\sigma_{DI}$ is of course a different concept than the implied volatility (which we denoted as $\sigma_I$). In this section we would like to make a simple, model independent, observation using $\sigma_{DI}$.  
 
Expanding (\ref{expansion1}) and comparing the appropriate terms we find one should take
\begin{equation} \label{frown}
\sigma_{DI}^2=\sigma^2\cdot \left(1-\frac{2}{c_m^2(1-\frac{x^2}{\sigma^2 t})}f(x,t,\sigma)+\ldots \right).
\end{equation}
Therefore, since $f(x,t,\sigma)$ is $\frown$ shaped, in general $\sigma_{DI}$ will also be $\frown$ shaped in variable $x$ (see Figure 1). 

\begin{figure}[h]
\centering
\includegraphics[width=80mm]{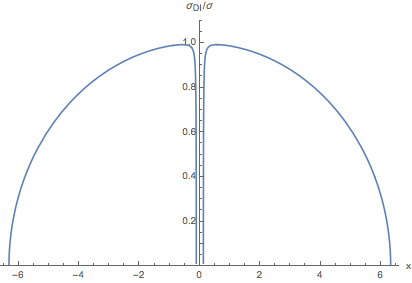}
\caption{Density implied volatility for $f=-1-10 x^4$ in $x$ variable for $c_m=3$, $t=0.5$, $\sigma=0.15$. The choice of $f$  is related to the actual result in Section 4 of this paper. Note the spike near the origin due to the singularity at $x^2 =\sigma^2 t$  in (\ref{frown}) which in turn appears due to locality in the market. In $S=e^x$ variable the plot would represent a skew.} 
\end{figure}

It is therefore a qualitative evidence of the fact that the volatility smile (which often takes the form of the skew) can be fairly easily explained by introducing causality to the Black-Scholes model. Note that the frown effect discussed above is generic i.e. it should appear in any financial model that introduces locality. A study  of a concrete realization of this idea is the main aim of this paper.

\section{Heat, Dirac and the telegraphers}
In this section we discuss relations between the heat, Dirac and the telegraphers equations in 1+1 dimensions. The material is well known to physicists and hence can be omitted if the reader is familiar with \cite{Goldstein, Kac, GJKS, JS, Orsingher1, Orsingher2}.

\subsection{Euclidean time}
 It is well known that the Schr\"odnger equation in imaginary time, for a free particle, results in the diffusion equation. This correspondence is only formal as the map $t \to -it$ has no physical reason. On the other hand, that map immediately suggests that if one would like to generalize the diffusion equation to the relativistic case one should use the Euclidean version of the Dirac equation. This reasoning can be captured in the following diagram
{\small
$$
\renewcommand{\arraystretch}{1.0}
\begin{array}{ccc} 
\hbox{Dirac equation} &\xrightarrow{v<<c}& \hbox{Schr\"odinger equation} \\
                \Big\downarrow\rlap{$t \to -it$}           &                                       &\Big\downarrow\rlap{$t \to -it$} \\
         \hbox{Euclidean Dirac equation} &   & \hbox{Heat equation}
\end{array}
$$
}
(where $v$ is a velocity, $c$ is the speed of light) when assumed that it closes.

However, by blindly using the Euclidean Dirac equation in this way, one loses understanding of the underlying stochastic process. For example, the object satisfying the Dirac equation is a spinor which in turn has many components (in our case,  1+1 dimensions,  the spinor has two real components). Therefore the interpretation of these components in terms of applications in finance, is not clear. However there exists a rigorous connection between stochastic processes and the Euclidean Dirac equation which follows form \cite{Kac} and  \cite{JS}, which we shall use.  

\subsection{Underling Poisson process}

Let us first start with the well known fact that the Wiener process, $W_t$, underlies the heat equation. We consider a particle on a line that follows  a simple random walk  (probability $1/2$ of going to the right or left). Given time $t$ define $p(x,t)$ as the probability density of a particle at point $x$. Then it can be easily shown that, in the limit of infinitesimal steps, $p(x,t)$ satisfies the heat equation. On the other hand, in the same limit considered, a simple random walk becomes the Brownian motion. In particular the coordinate of the particle is given by
\begin{equation} \label{wiener}
X(t) = X(0) +\sigma  \int_0^t dW_s
\end{equation}
 hence $W_t$ underlies the heat equation. This derivation is a bit sketchy (for a rigorous treatment see e.g. \cite{Shreve} or  \cite{Oksendal}) however it is very intuitive and  useful for further generalizations/modifications.

Let us now consider a stochastic process in which a particle travels along the line with constant velocity $v$ and changes the direction after time $\Delta t$ with probability $\lambda \Delta t$ where $\lambda$ is some constant. Now, consider two probability densities related to this process  $P_{+/-}(x,t)$, of a particle at time $t$, point $x$ with velocity to the right/left. It can be easily shown that, in the infinitesimal step limit,  $P_{+/-}(x,t)$ satisfy the telegraphers equation
\begin{equation} \label{telegraph}
\frac{2\lambda}{v^2}\partial_t P_{\pm} = \left(\partial_x^2 - \frac{1}{v^2}\partial_t^2\right)P_{\pm}.
\end{equation}
The same equation is satisfied for the probability density  $p(x,t):=P_+(x,t) + P_-(x,t)$ (a particle at time $t$, point $x$, any velocity) and the flow density $w(x,t):=P_+(x,t) - P_-(x,t)$.

On the other hand, in the limit $\Delta t \to 0$ the coordinate of the particle is given by
\begin{equation}  \label{tele}
X(t) = X(0)+v \int_0^t (-1)^{N(s)} ds
\end{equation}
where $N(s)$ is the number of events of the homogeneous Poisson process, at time $s$. Therefore a stochastic process underling the telegrapher equation is the Poisson process \cite{Goldstein, Kac}. Moreover in the $v \to \infty$ limit one recovers the Wiener process. 

\subsection{1+1 Dirac equation}

As pointed out in \cite{JS}, telegraphers equations are in fact equivalent to Euclidean version of the Dirac equation in $1+1$ dimensions. This is obtained by writing the Dirac equation in $1+1$ dimensions, explicitly for the components of the wave function $\Psi= (\psi_+, \psi_-)^T$. Introducing  the Euclidean time $t_E = i t$ (consequently the Euclidean speed of light  $c_E = -i c$), and defining new spinor components $u_\pm(t_E,x) = e^{\frac{mc_E^2}{\hbar}t_E}\psi_\pm$ one finds that $u_\pm$ satisfy the telegraphers equations provided we make the following identification: $t \leftrightarrow t_E$, $P_\pm \leftrightarrow u_\pm$, $v \leftrightarrow c_E$, $\lambda \leftrightarrow  \frac{mc_E^2}{\hbar}$.

Therefore one may conclude that the diagram discussed in the beginning of this section is not just formal. The Euclidean versions of the Dirac equation can be derived from the underling Poisson  process - the components $\psi_\pm$ of the spinor $\Psi$ correspond to probability densities $P_\pm$ multiplied by the factor $e^{-\lambda t}$.  
 
 \subsection{Fundamental solution}
 As indicated in \cite{Kac} the telegraphers equation becomes the heat equation in the $v \to \infty$ limit while keeping $\lambda/v^2$ fixed. Therefore the solutions of the telegraphers equation should converge to the solutions of the heat equation in that limit. Since telegraphers equation is second order in time derivatives one needs to fix the function and the first order derivatives at (say) $t=0$. Setting
 $$
 p(x,t)=\delta(x), \ \ \ \ \partial_t p(x,t) = 0 \ \ \ \ \hbox{for} \ \ \ \ t=0
 $$
 one can prove that the solution is \cite{Goldstein, Orsingher1, Orsingher2}
 $$
 p(x,t)= \frac{e^{-\lambda t}}{2v}\left[\delta(|x|-vt)+\lambda G(x,t) +  \partial_t G(x,t)\right],
 $$
\begin{equation} \label{solution}
G(x,t) = \begin{cases} I_0\left(\frac{\lambda}{v}\sqrt{v^2t^2-x^2}\right), & \mbox{for} |x|\le vt \\ 0, & \mbox{otherwise} \end{cases}
\end{equation}
where $I_0(z)$ is the order zero, modified Bessel function of the first kind. As shown in \cite{Orsingher2}, this solution indeed converges to the fundamental solution of the heat equation in the $v \to \infty$ limit.
 
\section{Generalizing  Black-Scholes}
Ideally one would like to use the Poisson process and its relation to the Wiener process (c.p. previous section) to derive the generalization of the Black-Scholes equation, using the standard hedging argument. Comparing the corresponding stochastic processes (\ref{wiener}) and (\ref{tele}) it seems reasonable to assume that a good starting point for the process describing the underlying asset $S(t)$ would be
\begin{equation} \label{rgbm}
\frac{dS_t}{S_t} = \mu dt + c_m(-1)^{N_t}dt,
\end{equation}
where we replaced $v$ with the maximal log-market velocity $c_m$. In the $c_m \to \infty$ limit, with $c_m/\sqrt{\lambda} = \sigma$ the term $c_m(-1)^{N_t}dt$ can be replaced by $\sigma dW_t$ (in a sense that the process (\ref{tele}) converges to (\ref{wiener}))  and one recovers the geometric Brownian motion. However it does not seem clear what is the analog of the Ito lemma for a process like (\ref{rgbm}) and we leave this problem for future investigations. Note that such counterpart of the Ito lemma would allow us to perform the standard hedging argument by designing the portfolio $\Pi=V(S,t)-\Delta S(t)$, i.e buying an option $V$ and selling $\Delta$ amount of stocks. Then, applying the generalized-Ito lemma for $d\Pi$, and requiring risk free portfolio, would result in a generalization of the Black-Scholes PDE.

In this section we will use a different route to arrive at the  option prices in the "relativistic" case.  We shall take advantage of the fact that under the risk neutral measure the price of the European option is given by the expectation value of the discounted payoffs. Denoting the payoff function by
$$
\mathcal{P}(S) :=   \begin{cases} \max(S-K,0), & \mbox{for a call option} \\ \max(K-S,0), & \mbox{for a put option}  \end{cases}
$$
the price of the European call/put will be
\begin{equation} \label{vanilla}
V(S,t)= e^{-r(T-t)}\int_0^\infty p(X(S'),T-t) \mathcal{P}(S')\frac{dS'}{S'},
\end{equation}
$$
X(S')= \ln \frac{S}{S'} + \left(r-\frac{1}{2}\sigma^2\right) (T-t),
$$
where $r$ is the interest rate, $p(x,\tau)$ is the risk neutral probability density. If $p(x,\tau)$ was given by the fundamental solution of the heat equation then (\ref{vanilla}) would give us the Black-Scholes formula for puts and calls. In our case $p(x,\tau)$ is a more complicated expression in terms of modified Bessel function (\ref{solution}).

\subsection{$1/c_m$ expansion}
Since in the $c_m \to \infty $ limit the exact solution (\ref{solution}) becomes the normal distribution, it is instructive to see what are the $1/c_m$ corrections before the limit is performed.

Following \cite{Orsingher2} we observe that in the large $c_m$ limit the argument of the Bessel function $I_0(\cdot)$ in  (\ref{solution}) is large, hence we can take advantage of the asymptotic expansion \cite{Abramowitz}
\begin{equation} \label{exp1}
I_0(z) = \frac{e^{z}}{\sqrt{2\pi z}}\left(1+\frac{1}{8z}+ \ldots \right), \ \ \ \ z>>1.
\end{equation}
The argument $z$ in our case can also be expanded as
\begin{equation}  \label{exp2}
z= \frac{\lambda}{c_m}\sqrt{c_m^2t^2-x^2}= \lambda t - \frac{\lambda x^2}{2c_m^2 t} - \frac{\lambda x^4}{8c_m^4 t^3 }+ \ldots \ \  .
\end{equation}
Note that since we have $\lambda=c_m^2/\sigma^2$, all the terms in (\ref{exp1}) and (\ref{exp2}) are necessary to capture all the $1/c_m^2$ contributions. On the other hand, to prove that $p(x,\tau)$ converges to the normal distribution, as it is done in \cite{Orsingher2}, one does not need the $1/8z$ term in (\ref{exp1}) and the $x^4$ term in (\ref{exp2}). Substituting (\ref{exp1}) and (\ref{exp2}) to (\ref{solution}) and using  $\lambda=c_m^2/\sigma^2$ we find that the solution (\ref{solution}) resolves as
\begin{equation} \label{exp3}
p(x,\tau) = \frac{e^{-\frac{x^2}{2\sigma^2 \tau}}}{\sqrt{2\pi \sigma^2 \tau}}\left(1+\frac{1}{c_m^2}f(x,\tau)+\ldots \right),
\end{equation}
\begin{equation} \label{corr}
f(x,\tau):= -\frac{\sigma^2}{8 \tau}+\frac{x^2}{2\tau^2}-\frac{x^4}{8\sigma^2 \tau^3}.
\end{equation}

\subsection{A crosscheck}
An independent way to verify (\ref{corr}) is to start with the telegraphers equation (\ref{telegraph}) and search for the solutions of the form of (\ref{exp3}).
Substituting (\ref{corr}) to (\ref{exp3}) we verify that the result satisfies the telegrapher equation up to the terms of order $1/c_m^2$ - as expected.

A more systematic way to see that is as follows. Using only the expansion (\ref{exp3}) we find that the telegraphers equation implies
$$
\frac{1}{2}\tau^2 \sigma^2 \partial_x^2f -\tau x\partial_xf -\tau^2 \partial_\tau f- \frac{x^4}{8\sigma^2\tau^2}+\frac{3x^2}{4\tau}-\frac{3\sigma^2}{8}=0
$$
where we neglected the terms of order $1/c_m^4$ and smaller. Now, we observe that the substitution $f(x,\tau)=w(\xi)/\tau$, $\xi=x^2/\tau$ results in an ordinary differential equation for $w(\xi)$
\begin{equation} \label{corr1}
2\sigma^2 \xi w''(\xi)+(\sigma^2-\xi)w'(\xi)+w(\xi)+\frac{3\xi}{4}-\frac{\xi^2}{8\sigma^2}-\frac{3\sigma^2}{8}=0
\end{equation}
for which the most general, quadratic in $\xi$, solution is
$$
w(\xi) = -\frac{1}{8\sigma^2}\xi^2+\left(\frac{3}{8}-\frac{a}{\sigma^2}\right)\xi +a, \ \ \ \ a\in \mathbb{R}.
$$
Taking $a=-\sigma^2/8$ we see that $w(x^2/\tau)/\tau$ coincides with (\ref{corr}). Therefore we have shown that the $1/c_m^2$ corrections (\ref{corr}) are consistent with the expansion (\ref{exp3})

The general solution of (\ref{corr1}) can be obtained by finding the general solution of the corresponding homogeneous equation and adding it to the special solution $w(\xi)$. The result is
{\tiny
$$
w_{gen}(\xi)=\frac{\xi-\sigma^2}{2\sigma}c_1 + \left[ \frac{2\sqrt{\xi}}{\pi}+\sqrt{\frac{2}{\pi}}\erfi \left(\frac{\sqrt{\xi}}{\sqrt{2}\sigma}\right)\left(\sigma-\frac{\xi}{\sigma}\right)\right]c_2 + w(\xi)
$$  
}
where $\erfi(x)$ is the imaginary error function $\erfi(x):=-i \erf(ix)$. The fact that the above solution is not unique is reasonable since we did not specify the boundary conditions for $w(\xi)$.

\subsection{Black-Scholes formula with $1/c_m^2$ corrections}
A complete treatment of the problem requires calculating the exact value of e.g. the call, which we shall do now.  Substituting (\ref{exp3}) and (\ref{corr}) to (\ref{vanilla}) one finds that the call option is
{\small
\begin{equation}  \label{integral}
V(S,\tau)=\frac{K e^{-r \tau}}{\sqrt{2\pi \sigma^2 \tau}} \int_0^{y_{max}} (e^y-1)e^{-\frac{(x-y)^2}{2\sigma^2\tau}}\left[1-\frac{1}{c_m^2}f(y,\tau)\right]dy,
\end{equation}
$$
x= \ln S/K + \left(r-\frac{1}{2}\sigma^2\right) \tau
$$
}
where we changed the integration variables for convenience. The upper integration limit $y_{max}$ is given implicitly by $f(y_{max},\tau)=c_m^2$ which has four solutions,  however only one of them is always real and positive
$$
y_{max}= \sqrt{2\sigma^2 \tau +\sigma \tau \sqrt{3\sigma^2+8c_m^2 \tau} }.
$$
In the limit $c _m\to \infty$ we have $y_{max}\to \infty$ and the integral (\ref{integral}) results in the Black-Scholes formula. For finite $c_m$ the integration is more complicated.
Because of the exponential damping of the integrand we will approximate the integral by assuming that $y_{max}=\infty$. By dong so we introduce a negligible error compared to the $1/c_m^2$ corrections that are already in the integrand.  However now the integral is  elementary since $f(y,\tau)$ is a (quartic) polynomial in $y$.  The final result is relatively simple in terms of standard $d_1$, $d_2$ parameters
\begin{equation} \label{bscorr}
V(S,\tau)=S N(d_1)-K e^{-rt}N(d_2) +\frac{1}{c_m^2}v,
\end{equation}
$$
d_1=\frac{\sigma^2 \tau+x}{\sigma \sqrt{\tau}}, \ \ \ \ d_2 = \frac{x}{\sigma \sqrt{\tau}}
$$
where
{\tiny
$$
v=-\frac{\sigma^2}{8\tau} [S M(d_1)-K e^{-rt}M(d_2)  ]
-\frac{S \sigma^2}{8\sqrt{2\pi \tau}} e^{-\frac{d_1^2}{2}}\left(1+\frac{3}{2}d_1^2+\frac{3}{2}d_2^2-\frac{1}{2}\sigma^2 \tau \right)
$$
}
where
$$
M(z):=N(z)z^2(z^2+2), \ \ \ \ N(z)=\frac{1}{\sqrt{2\pi}} \int_{-\infty}^z e^{-t^2/2}dt.
$$
Having derived the $1/c_m^2$ corrections to the Black-Scholes formula we are now in the position to find the corresponding implied volatility. To this end we make a similar analysis as in Section 2. We examine how the Black-Scholes formula changes when $\sigma \to \sigma \cdot (1 + s)$ where $s$ is small. The $d_1$ and $d_2$ parameters become
$$
d_1 \to d_1 +s \left(\sigma \sqrt{\tau} - \frac{x}{\sigma\sqrt{\tau}} \right), \ \ \ \ d_2 \to d_2 -\frac{ s x}{\sigma \sqrt{\tau}}
$$
and hence the Black-Scholes formula $V_{BS}=S N(d_1)-K e^{-r\tau}N(d_2) $ is
$$
V_{BS}\to V_{BS} + s \bar{v},
$$
\begin{equation}  \label{bsvar}
\bar{v} =  \frac{1}{\sigma\sqrt{2\pi \tau}} \left[ S (\sigma^2 \tau -x) \sigma e^{-\frac{d_1^2}{2}} + K x e^{-r \tau-\frac{d_2^2}{2}}   \right].
\end{equation}
Comparing (\ref{bsvar}) with (\ref{bscorr}) we find that $s=v/\bar{v}c_m^2$ and hence  the implied volatility is
\begin{equation} \label{implVol}
\sigma_I = \sigma \cdot \left(1+ \frac{v}{c_m^2\bar{v}}\right).
\end{equation}
This result is plotted in Figure 2.
\begin{figure}[h]
\centering
\includegraphics[width=70mm]{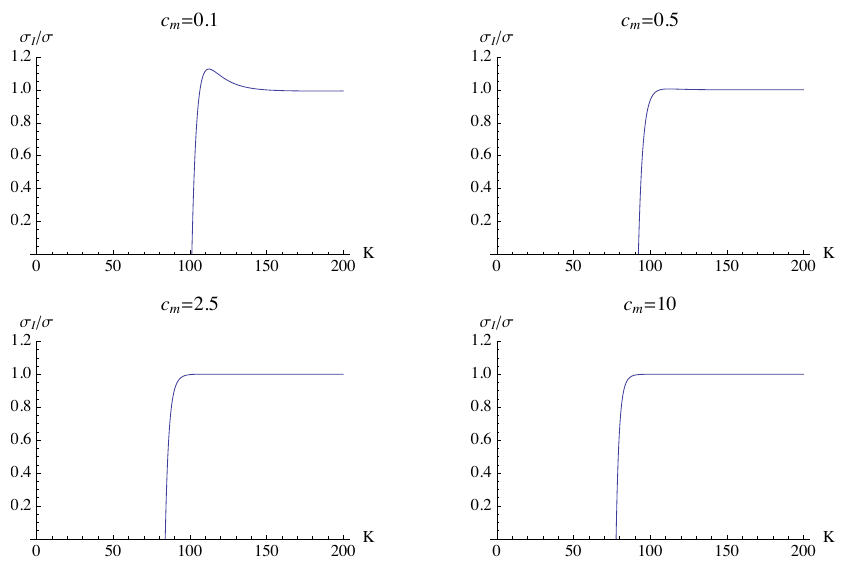}
\caption{$1/c_m^2$ corrections of the implied volatility given by (\ref{bsvar}), for $S=100$, $T-t=0.5$, $r=0.05$, $\sigma=0.15$. }
\end{figure}

\section{Summary and Outlook}

Relativistic extensions of the Black-Scholes model seem very natural, considering future development of high frequency trading. However the physical bound on the maximal speed of the asset is, to our understanding, still too high to give noticeable effects in the market. On the other hand, as we argued in the introduction, the effective maximal speed of log-returns, $c_m$, is much smaller due to the "resistance" of the market - an analogous phenomena appears in some physical situations.  Therefore relativistic extensions with such effective velocity, instead of the real one, seem reasonable.

In this paper we  considered a certain relativistic extension of the Black-Scholes model, based on the observation that the Black-Scholes equation, in particular coordinates, becomes a heat equation. The latter is clearly non relativistic and therefore it is a good starting point for relativistic extensions. The stochastic process behind the heat equation is a Brownian motion, which implies that an appropriate extension should be related to a process such that in  the $c_m \to \infty$ limit the Wiener process is recovered. A very well known process which satisfies this condition is the telegrapher process. Not only does it converge to the Wiener process in the above limit but also, it incorporates the features of relativity in a very clever way: the system of PDEs describing the probability densities of the telegrapher process is equivalent to the Euclidean version of the Dirac equation in $1+1$ dimensions. Therefore it provides an extremely elegant framework. Our most important finance-related conclusion based on these remarks is that the geometric Brownian motion should be replaced by its relativistic counterpart (\ref{rgbm})
$$
dS_t/S_t = \mu dt + c_m (-1)^{N(t)}dt,
$$ 
where $N(t)$ is the number of events in the homogeneous Poisson process with rate parameter $\lambda$. This SDE becomes the geometric Brownian motion with volatility $\sigma$ when the $c_m \to \infty$ limit is performed (keeping $\lambda = c_m^2/\sigma^2$). It is not an Ito process and therefore one cannot use the Ito lemma to derive the corresponding equation for a derivative instrument. We circumvent this problem by using the derivative pricing formula in the risk neutral measure, replacing the Gaussian probability distribution (used in the Black-Scholes model) by its relativistic counterpart. In that case the pricing formula is given by Eq. (\ref{vanilla}) with risk neutral probability (\ref{solution}). We  evaluated the $1/c_m^2$ corrections to the Black-Scholes formula, using Eq. (\ref{vanilla}), and found that the corresponding implied volatility resembles the frown shape.

There are some direction where one can improve our results and the model itself. One is to perform  thorough Monte Carlo simulations based on the SDE (\ref{rgbm}) which could then be compared  with (\ref{bsvar}). Second, it would be very interesting to derive a counterpart of the Ito lemma for the process (\ref{rgbm}) as it could be used to derive the pricing PDE from first principles. 

Lastly one could generalize the process (\ref{rgbm}) by using non-constant effective velocity $c_m$ (because it is effective  there is a priori no reason to assume that it is constant).  Clearly one could also consider a stochastic process for $c_m$ (e.g. some mean-reverting process) 
$$
dc_m^2 = \alpha(c_m,t)dt + \beta(c_m,t) dW_t
$$
which, together with (\ref{rgbm}) and the constraint $c_m^2 = \lambda \sigma^2$, would result in a certain generalization of stochastic volatility models.  The randomness of volatility would then be explained by the randomness of $c_m$ since  $d\sigma^2 = \lambda^{-1} d c_m^2$. Furthermore one can also consider a non-homogeneous Poisson process (i.e. with non constant $\lambda$)  
therefore adding one more degree of freedom to the model.

\section{Acknowledgements}
I would like to thank G. Araujo, T. Banerjee, N. Butt, A. Gomez, M. Kust, M. Maurette and S. Mercuri for reading the manuscript and comments. I also thank M. Ku\'zniak for email correspondence and  B. Trzetrzelewska for encouragement. This work is supported in part by the NCN grant: UMO-2016/21/B/ST2/01492.

\end{document}